\documentclass[prb,preprint]{revtex4-1} 
\usepackage{amsmath}  % needed for \tfrac, \bmatrix, etc.
\usepackage{amsfonts} % needed for bold Greek, Fraktur, and blackboard bold
\usepackage{graphicx} % needed for figures
\usepackage{adjustbox}
\usepackage{relsize}
\usepackage{comment}
\usepackage{rotating}
 
\begin{document}

% Be sure to use the \title, \author, \affiliation, and \abstract macroshttps://www.overleaf.com/project/63cec1dc13e17a68a85cf119
% to format your title page.  Don't use lower-level macros to manually
% adjust the fonts and centering.

\title{Another reason why normalized gain should continue to be used to analyze concept inventories (and estimate learning rates)}
% In a long title you can use \\ to force a line break at a certain location.

%When submitting the manuscript for review, do not include the author's name or institution
\author{Jairo Navarrete}
\author{Valentina Giaconi}
% \author statement.
\affiliation{Instituto de Ciencias de la Educaci\'on, Universidad de O'Higgins, Rancagua, Chile}
% Please provide a full mailing address here.

\author{Gonzalo Contador}
%\email{ajp@dickinson.edu}
\affiliation{Departmento de Matem\'atica, Universidad T\'ecnica Federico Santa Mar\'ia, Valpara\'iso, Chile}

\author{Mariano Vazquez}
\affiliation{Departmento de Ingenier\'ia Matem\'atica, Universidad de Chile, Santiago, Chile}

% See the REVTeX documentation for more examples of author and affiliation lists.

%\date{\today}

%\begin{abstract}
%Manuscripts intended for the Notes and Discussions section %should be considerably shorter, typically 1000 to 3000 words.
%\end{abstract}
% AJP requires an abstract for all regular article submissions.
% Abstracts are optional for submissions to the "Notes and Discussions" section.
\begin{abstract}
A transformation called normalized gain (ngain) has been acknowledged as one of the most common measures of knowledge growth in pretest-posttest contexts in physics education research. Recent studies in math education have shown that ngains can also be applied to assess learners' ability to acquire unfamiliar knowledge, that is, to estimate their ``learning rate''. This quantity is estimated from learning data through two well-known methods: computing the {\it average ngain of the group} or computing the {\it ngain of the average learner}. These two methods commonly yield different results, and prior research has concluded that the difference between them is associated with a pretest-ngains correlation. Such a correlation would suggest a bias of this learning measurement because it implies its favoring of certain subgroups of students according to their performance in pretest measurements. The present study analyzes these two estimation methods by drawing on statistical models. Our results show that the two estimation methods are equivalent when no measurement errors exist. In contrast, when there are measurement errors, the first method provides a biased estimator, whereas the second one provides an unbiased estimator. Furthermore, these measurement errors induce a spurious correlation between the pretest and ngain scores. Our results seem consistent with prior research, except they show that measurement errors in pretest and posttest scores are the source of a spurious pretest-ngain correlation. Consequently, estimating learning rates might effectively provide unbiased estimates of knowledge change that control for the effect of prior knowledge even in the presence of pretest-ngain correlations.
\end{abstract}
\maketitle % title page is now complete

\section{Introduction} % Section titles are automatically converted to all-caps.
% Section numbering is automatic.

%Possible emails for Lei BAO:
%gvmr5y6@163.com
%lbao@mps.ohio-state.edu
%web: https://www.asc.ohio-state.edu/bao.15/test1/Research.htm

Learning data from pretest-posttest designs can be transformed into a unidimensional measure called normalized gain (ngain) that estimates learning. In physics education research, ngains were popularized by Richard Hake in a study from 1998 \cite{hake_interactive-engagement_1998}. Concept Inventories are widely used in this field to test the efficacy of various instructional methods, including assessment batteries such as the Force Concept Inventory (FCI), Force and Motion Concept Evaluation, and the Conceptual Survey of Electricity/Magnetism. In Hake's study, FCI data from 62 courses with 6542 students was analyzed to provide compelling evidence of the superiority of interactive engagement courses in physics against those taught using traditional methods. Hake introduced this score as ``a rough measure of the effectiveness of a course in promoting conceptual understanding.'' Since Hake published his 1998 study, ngains have been widely used in the teaching of the physics research community and acknowledged as the most common measurement of gains used in physics education research \cite{nissen_comparison_2018}. This measure has been previously applied to individuals \cite{gery1972does} by considering pretest scores, post-test scores, and the maximum possible score (M). Its computation transforms two learning variables (pretest and post-test scores) into a uni-variate measure. When $M$ is the maximum achievable in these assessments, the ngain of each learner is defined \cite{gery1972does} as follows:
\begin{equation}
\label{eq:ng}
ng = \frac{{postest}-{pretest}}{M-pretest} 
\end{equation}

A recent study by Nissen J. et al.\cite{nissen_comparison_2018} have argued ``that [ngains] is biased in favor of high pretest populations'' as they observed non-zero pretest-ngains correlations, thus implying that ngains favor students according to their prior knowledge. A response to this criticism argued a bias of the study due to omitting control variables in the analysis\cite{coletta_why_2020}. In this work, we will show that even when ngains were a fair measure for learning (meaning that does not favor any subgroup of students), a small ngains-pretest correlation would be induced due to measurement errors in pretest and posttest scores. 

Recent research in mathematics education has presented empirical evidence that this measure can also be applied to assess the learning of mathematics\cite{navarrete_learning_2024}. Through the development of a statistical model---called the Learning Rate Model (LR-Model)---the study proposes a construct called the ``Learning Rate'', defined as the learner's ability to acquire unfamiliar knowledge under particular instruction conditions. In this context, the ngains transformation allows us to estimate the learning rate of students. The referred study presents ngains as a correction of raw gains (gain = posttest-pretest) and inferred its mathematical properties to afterward associate them with learning phenomena. For example, the mathematical theory shows that a positive learning rate generates a negative pretest-gain correlation, which is then associated with learning phenomena as follows:
\begin{quote} 
This theoretical result can be linked to an old puzzle in educational measurement. Early and contemporary research has reported negative pretest-gain correlations in learning studies \cite{cronbach_how_1970, hake_interactive-engagement_1998, hake_relationship_2002, oconnor_jr_extending_1972}. Intuitively,
these negative correlations are unexpected since smaller gains correspond to learners with more considerable prior knowledge, and thus, early researchers treated these negative correlations as anomalies. For example, Thorndike (1924) explained such a negative bias was a spurious phenomenon due to measurement errors in the pretest (see also Thorndike, 1966). Similarly, Thomson (1924, 1925) and Zieve (1940) suggested analytic procedures to adjust for this negative bias. In contrast, the Learning Rate Model predicts these negative correlations, provides a rationale for them, and quantifies them by explaining its dependency on the learning rate F \cite{navarrete_learning_2024}. (Navarrete, 2024)
\end{quote}

Assessing the learning rate of a group of learners is crucial for research, and thus, prior work has discussed estimating this quantity from learning data in two manners \cite{bao_theoretical_2006, marx_normalized_2007}. The first computes the \emph{average ngain of the group} of learners, whereas the second computes the \emph{ngain of the average learner}. These two estimators are computed as follows:
\begin{align}
\label{eq:average_ng}
\overline{ng} &= \frac{1}{N}\sum {ng}_i \\
\label{eq:ng_of_average_learner}
\widehat{ng} &= \dfrac{\frac{1}{N}\sum{postest_i}-\frac{1}{N}\sum{pretest_i}}{M-\frac{1}{N}\sum{pretest_i}}
\end{align}

A prior study argued that the difference between these two estimators provides valuable information about the learning process\cite{bao_theoretical_2006}. By drawing on analyses of idealized situations, Lei Bao concluded that the difference between them gives insight into the group's learning process. More specifically, the study states:
\begin{quote}
If [$\overline{ng}$] is greater than [$\widehat{ng}$], we can infer that students with low pre-test scores tend to have either smaller or similar score improvement than students with high pre-test scores. On the other hand, if [$\overline{ng}$] is smaller than [$\widehat{ng}$], students with low pre-test scores tend to have larger score improvements than students with high pre-test scores. (Lei Bao, 2006)
\end{quote}

The above quote conveys a relationship between these two estimators' subtraction and the correlation pretest-ngains. For the sake of conciseness, this study summarizes the quoted relationship by the following statement:
\begin{align}
\label{baoStatement}
\overline{ng} - \widehat{ng} \geq 0 \textbf{ \ exactly when }\text{ pretest-ngains correlation is positive (or zero).}
\end{align}
The pretest-ngains correlation has caught the attention of researchers in the field of learning measurement as it pinpoints a fundamental problem in measuring learning: it might indicate an inability of the measure to compare changes at different levels of pretest performance. In other words, such a correlation suggests a bias favoring certain groups of learners over others. Indeed, early research has proposed that ``optimal measures of knowledge change'' should not correlate with pretest scores \cite{oconnor_jr_extending_1972}. There are mixed views about the pretest-ngains correlation. On the one hand, empirical evidence has shown that ngains and pretest scores are uncorrelated, thus supporting that the ngains transformation effectively measures learning by controlling the influence of prior knowledge \cite{hake_interactive-engagement_1998, hake_relationship_2002, coletta_interpreting_2005, coletta_interpreting_2007, coletta_why_2020, navarrete_learning_2024}. On the other hand, other studies have observed small correlations between pretest scores and ngain scores \cite{nissen_comparison_2018, bao_theoretical_2006}, thus arguing that ngain scores do not control for prior scores influence. These latter arguments have been rebutted by observing that the performed analysis is biased for ignoring crucial control variables (e.g., reasoning abilities)\cite{coletta_why_2020}.

In this context, notice that Bao's conclusion (statement \ref{baoStatement}) is crucial as it translates the difference between these two estimators into the existence of a pretest-ngains correlation, and thus, the inability of ngains to discount the influence of prior knowledge. In real settings, the group's average ngain $\overline{ng}$ generally differs from the average learner's ngain $\widehat{ng}$, thus rendering ngains as a biased measure of student growth. The main goal of the present work is to provide a second viewpoint where, even considering that Bao's statement (\ref{baoStatement}) holds, ngains are not rendered as a biased measure. To this aim, we will show below that measurement errors in pretest and posttest scores induce a spurious pretest-ngain correlation even when prior knowledge (pretest) and learning rates (ngain) are statistically independent.

The distinction between the two viewpoints is crucial. The first one yields inferences about the learning phenomenon and suggests a bias of the growth measure (ngain). The second viewpoint neither yields inferences nor suggestions. The difference $(\overline{ng}-\widehat{ng})$ appears often in practice, and thus, associating it with measurement errors rather than with a biased measurement is crucial. Hence, the main contribution of this study is to yield a different rationale for Bao's statement (\ref{baoStatement}) while providing theoretical arguments that support ngain scores as a measure of learning that effectively controls for the influence of prior knowledge.

 A second contribution of this study is to provide a framework for studying the behavior of measurement errors when estimating learning rates. By modifying the statistical model introduced in a prior study\cite{navarrete_learning_2024}, we will analyze the behavior of the two learning rate estimators ($\overline{ng}$ and $\widehat{ng}$) in the presence of measurement errors. This framework might be useful as measurement errors are unavoidable in learning measurement.

The current work has the following structure. The second section presents a statistical model recently developed called the Learning Rate Model (LR-Model)\cite{navarrete_learning_2024}. We will adopt the notation of such a model as it facilitates stating the problem accurately. The third section yields theoretical developments regarding the case without measurement errors. The fourth section modifies the LR-Model to address the study of measurement errors. The fifth section yields theoretical developments regarding the case of measurement errors. When measurement errors are considered, our results are consistent with Bao's results, thus showing support for statement (\ref{baoStatement}). We will also show that measurement errors induce a spurious ngain-pretest correlation. The sixth section yields statistical simulations to support our theoretical developments. The last section will discuss the distinction between the two interpretations of statement (\ref{baoStatement}) under analysis. Finally, we will provide some suggestions for estimating learning rates using ngains. 

\section{The Learning Rate Model } \label{LR_M}
The Learning Rate Model (L-R Model) was recently introduced\cite{navarrete_learning_2024} as a theoretical construct underlying ngain scores (equation \eqref{eq:ng}). The model assumes that the knowledge delivered by instruction is bounded and that the learner's pretest $X$ and posttest $Y$ have no measurement errors. Also, $0 \leq X \leq Y \leq 1$ and $X < 1$ (with probability $1$). Without loss of generality, it is assumed that the theoretical maximum score $M$ in \eqref{eq:ng} is standardized to 1 (100\% achievement). The referred study\cite{navarrete_learning_2024} presents theoretical arguments and empirical evidence supporting that the final status of knowledge $Y$ results from two independent random variables: the initial status of knowledge $X$ and the learning rate $F$. In addition to the assumption that $X$ and $F$ are independent, the other key hypothesis of the model is that the raw gains of knowledge $G = Y-X$ are proportional to both the amount of unfamiliar knowledge $(1-X)$ encountered by the learner during the learning process and the learning rate $F$. In other words, $G = (1-X)F$. More precisely, the LR-Model states that the relation between these variables is the following one:
\begin{align}
\label{eq:posttest}
    Y &= X + (1-X)F,
\end{align}

The intended interpretation is that the learning rate $F$ represents the learner's ability to acquire unfamiliar knowledge under a specific learning environment. The learning rate $F$ is a theoretical construct estimated empirically by either $\overline{ng}$ or $\widehat{ng}$. The expected value of $F$ (\ref{eq:meanF}) can be obtained by taking the expected value to both sides of equation (\ref{eq:posttest}).
\begin{align}
\label{eq:defF}
F &= \dfrac{Y-X}{1-X}\\
\label{eq:meanF}    
\mu_F &= \dfrac{\mu_Y-\mu_X}{1-\mu_X}
\end{align}
The learning rate $F$ is a random variable, and thus, its mean can be estimated from learning data. We can assume two random variables to model the learning phenomenon, namely, let $\{(X_1, F_1), (X_2, F_2),..., (X_n, F_n),\}$ be $n$ pairs of independent and identically distributed random variables. Now let $\{Y_1, Y_2, ..., Y_n\}$ following equation (\ref{eq:posttest}) to determine learners' posterior performance. In such case, we can define the two following random variables as estimators of the mean of the learning rate $\mu_F$: 
\begin{align}
\label{eq:barf}
\overline{F} &= \frac{1}{n}\sum \frac{Y_i-X_i}{1-X_i}  \\
\label{eq:hatf}
\widehat{F} &= \dfrac{\frac{1}{n}\sum{Y_i}-\frac{1}{n}\sum{X_i}}{1-\frac{1}{n}\sum{X_i}}
\end{align}
 Note that $\overline{F}$ and $\widehat{F}$ are theoretical counterparts of $\overline{ng}$ and $\widehat{ng}$, respectively. The next section shows that both estimators are unbiased when there are no measurement errors. 

\section{Error-less measurement assumption} \label{sec:noerror} 
Under the hypothesis that learning is accurately represented by equation (\ref{eq:posttest}) and that the pretest ($X$) and the posttest ($Y$) are observed without measurement errors, we aim to obtain two key results. The first is that $\overline{F}$ is an unbiased estimator of $\mu_F$. The second result is that $\widehat{F}$ is an asymptotically unbiased estimator for $\mu_F$. In other words, without measurement errors, equations (\ref{eq:barf}) and (\ref{eq:hatf}) determine unbiased estimators for $\mu_F$.

To analyze the estimator $\overline{F}$ (eq. \ref{eq:barf}), let us denote by $F_i$ to the $i$-th students' learning rate and compute the expected value on each side of the equality below. The result shows that $\overline{F}$ is an unbiased estimator of $\mu_F$. Furthermore, a direct application of the central limit theorem shows that $\overline{F}$ is also a consistent estimator for $\mu_F$.  
\begin{align*}
\overline{F}\  &= \frac{1}{n}\sum_{i=1}^n F_i \\
\mathbb{E}[\overline{F}] &= \frac{1}{n}\sum_{i=1}^n \mu_F \\
\mathbb{E}[\overline{F}] &= \mu_F
\end{align*}

To analyze the estimator $\widehat{F}$ (eq. \ref{eq:hatf}), we can use previous results\cite{navarrete_learning_2024} to compute the covariance matrix $COV(X,Y)$:

$$\Sigma=\begin{bmatrix} \sigma^2_X & \left(1-\mathbb{E}(F)\right) \sigma^2_X  \\ \left(1-\mathbb{E}(F)\right) \sigma^2_X & \sigma^2_Y \end{bmatrix}$$ 
and thus, an application of the central limit theorem indicates that 
  \begin{equation}
        \label{eq:meanasymp}
        \sqrt{n}\left(\begin{pmatrix}\Bar{X} \\ \Bar{Y} \end{pmatrix} - \begin{pmatrix}\mathbb{E}(X)\\ \mathbb{E}(Y) \end{pmatrix}\right) \to \mathcal{N} \left(\begin{pmatrix}0 \\ 0 \end{pmatrix},  \Sigma\right).
    \end{equation}
We can show that $\widehat{F}$ is an asymptotically unbiased estimator for $\mu_F$ by noticing that $\widehat{F}=f(\bar{X}, \bar{Y})$ where $f(x,y)=\frac{y-x}{1-x}$, and $f(\mu_X, \mu_Y) =\mu_F$ (eq. \ref{eq:meanF}). The delta method yields:  
    \begin{equation}
        \label{eq:fhatasymp}
        \sqrt{n}\left(\widehat{F}-\mu_F\right) \to \mathcal{N} \left(0,  \sigma^2_{\hat{F}}\right), \text{ where}
    \end{equation}
\[\sigma^2_{\hat{F}}= \dfrac{[\sigma^2_Y(1-\mu_X)^2-\sigma^2_X(1-\mu_Y)^2]}{(1-\mu_X)^{4}}\] 

The distributional result (eq. \ref{eq:fhatasymp}) implies that  $\widehat{F}$ is a consistent estimator for $\mu_F$ and thus asymptotically unbiased. Furthermore, as $F_i$ is independent of the sequence $X_1, \cdots, X_n$, the tower property implies that:
\begin{align*}
\mathbb{E}(\widehat{F}) &= \mathbb{E}\left( \frac{\frac{1}{n} \sum_i (Y_i-X_i)}{\frac{1}{n}\sum_i (1-X_i)}\right)\\
                        &= \mathbb{E}\left(\frac{\sum_i F_i(1-X_i)}{\sum_i (1-X_i)}\right)\\
                        &= \mathbb{E}\left(\mathbb{E}\left(\frac{F_1(1-X_1)}{\sum_i (1-X_i)}+\frac{F_2(1-X_2)}{\sum_i (1-X_i)}+...+\frac{F_n(1-X_n)}{\sum_i (1-X_i)}\biggr|X_1, \cdots ,X_n\right)\right)\\
                        &= \mathbb{E}\left(\frac{\mu_F(1-X_1)}{\sum_i (1-X_i)}+\frac{\mu_F(1-X_2)}{\sum_i (1-X_i)}+...+\frac{\mu_F(1-X_n)}{\sum_i (1-X_i)}\right)\\
                        &=\mu_F
\end{align*}
                      
As both estimators are unbiased, the relative efficiency of $\overline{F}$ with respect to $\widehat{F}$ will be given by their variance ratio. A straightforward result is that $\overline{F}$ is asymptotically more efficient than $\widehat{F}$ whenever
\begin{comment}
thus $\overline{F}$ will be asymptotically more efficient than $\widehat{F}$ whenever $\sigma^2_F(1-\mu_X)^4<\sigma^2_Y(1-\mu_X)^2-\sigma^2_X(1-\mu_Y)^2$. As $Y=X+F-FX$, one has that 
\begin{align*}
   \sigma^2_Y&= \sigma^2_X+\sigma^2_F+Var(FX)-2[Cov(X,FX) + Cov(F,FX)]\\
   &=\sigma^2_X+\sigma^2_F+\mathbb{E}(F^2)\mathbb{E}(X^2)-\mu_F^2\mu_X^2-2[\sigma^2_X\mu_F + \sigma^2_F\mu_X]\\
   &=\sigma^2_X(1-\mu_F)^2+\sigma^2_F(1-\mu_X)^2+ \sigma^2_X\sigma^2_F
\end{align*}

So  $\overline{F}$ will be asymptotically more efficient than $\widehat{F}$ whenever 
\begin{align}
\label{efficiencyCondition}
\left(\frac{1-\mu_Y}{1-\mu_X}\right)^2<\mathbb{E}((1-F)^2)=(1-\mu_F)^2+\sigma^2_F
\end{align}
\end{comment}
\begin{align}
\label{efficiencyConditionJairo}
\left(\frac{1-\mu_Y}{1-\mu_X}\right)^2 &< (1-\mu_F)^2+\sigma^2_F
\end{align}

To summarize, when there are no measurement errors, both estimators $\overline{ng}$ and $\widehat{ng}$ are unbiased. It is worth emphasizing, in this case, that any difference between them is due only to randomness.
% DEJE TODOS ESTOS DESARROLLOS ACA, VALENTINA Y JAIRO VEAN SI SE MUEVEN AL SUPLEMENTO O AL APENDICE DE ESTE DOC. sI QUIEREN TB DESMENUZAMOS/INTERPRETAMOS UN POCO ESTO EN LA PROXIMA REU

\section{The LR-Model and Measurement Errors}  
In Psychometrics, a common assumption is considering that a test measures certain true variable plus an additive error \cite{shultz2020measurement}. Hence, here we define a useful random variable analogous to equation (\ref{eq:defF}) for including measurement errors for pretest and posttest instruments. We denote these errors by $\epsilon_X$ and $\epsilon_Y$, respectively, and assume they have zero means and are independent of every other random variable. It is important to emphasize that external observers of the learning phenomenon have access only to the observed noisy values of pretest ($X^*=X+\epsilon_X$) and posttest ($Y^*=Y+\epsilon_Y$) scores. Consequently, they have access only to the noisy learning rate variable $F^*$ defined as follows:
 \begin{equation}
        \label{eq:f_0}
        F^*=\frac{Y^*-X^*}{1-X^*}
\end{equation}
%=\frac{Y-X+\epsilon_Y-\epsilon_X}{1-X-\epsilon_X}.

Let us compute the expected value of this random variable. By taking expected value to both sides of the equality $F^*(1-X^*) = Y^*-X^*$
to obtain that $\mathbb{E}(Y-X) = \mathbb{E}(F^*)\mathbb{E}(1-X)+Cov(F^*, 1-X^*)$. By solving for $\mathbb{E}(F^*)$ we obtain:
\begin{align}
    \label{eq:biasFstar}
    \mathbb{E}(F^*) &=\mu_F -\frac{Cov(F^*, 1-X^*)}{\mathbb{E}(1-X)} \\ 
    &=\mu_F -Cov\left(F^*, \frac{1-X^*}{\mathbb{E}(1-X)} \right)\\ 
     &=\mu_F +\frac{Cov(F^*, X^*)}{\mathbb{E}(1-X)} \\
     &= \mu_F +\frac{Cov(F^*, X^*)}{1-\mu_X}
\end{align}
Although the random variables $X$ and $F$ remain independent (i.e., $Cov(X, F) = 0$), it does neither imply that $Cov(X, F^*) = 0$ nor $Cov(X^*, F^*) = 0$. Since $F^*$ is defined from $X$ and $\epsilon_X$, the random variable $F^*$ statistically depends on $X$ xor $X^*$. Hence, $F$ and $F^*$ have similar but not identical means. 

As before, let us define two estimators for the mean of learning rates. Based on $\{(X_{1}, F_{1}), (X_{2}, F_{2}),\cdots, (X_{n}, F_{n})\}$  $n$ pairs of independent and identically distributed random variables, we define $\{Y_{1}, Y_{2}, \cdots, Y_{n}\}$ by following equation \eqref{eq:posttest}. Let us also consider the pairs of errors $\{ (\epsilon_{X_1}, \epsilon_{Y_1}), (\epsilon_{X_2}, \epsilon_{Y_2}), \cdots, (\epsilon_{X_n}, \epsilon_{Y_n})\}$ associated to observations $\{(X^*_{1}, Y^*_{1}), (X^*_{2}, Y^*_{2}),\cdots, (X^*_{n}, Y^*_{n})\}$. The family of observed learning rates $\{F^*_{1}, F^*_{2}, ..., F^*_{n}\}$ is determined by equation (\ref{eq:f_0}). Let us define the following estimators of $\mu_F$:

\begin{align}
\label{eq:barf0}
\overline{F^*} &= \frac{1}{n}\sum_{i=1}^n \frac{Y^*_{i}-X^*_{i}}{1-X^*_{i}} =\frac{1}{n}\sum_{i=1}^n F^*_i \\
\label{eq:hatf0}
\widehat{F^*} &= \dfrac{\frac{1}{n}\sum_{i=1}^n{Y^*_{i}}-\frac{1}{n}\sum_{i=1}^n{X^*_{i}}}{1-\frac{1}{n}\sum_{i=1}^n{X^*_{i}}}
\end{align}
%=\frac{\overline{Y^*}-\overline{X^*}}{1-\overline{X^*}}
As before, the definitions of $\overline{F^*}$ and $\widehat{F^*}$ are theoretical counterparts of $\overline{ng}$ and $\widehat{ng}$.  In the following section, we will show that $\widehat{F}$ is an asymptotically unbiased estimator of $\mu_F$ and that $\overline{F*}$ is a biased estimator of $\mu_F$. As a byproduct of this latter result, we will show that Bao's statement (\ref{baoStatement}) holds. 

\section{Results under measurement errors}
\begin{comment}
    Here, we show that $\overline{F^*}$ carries a systematic bias as an estimator of $\mu_F$, related to the error in data measurement, whereas $\widehat{F^*}$ is an asymptotically unbiased estimator of $\mu_F$. $\widehat{F^*}$ will also be biased based on a finite sample, but its bias will turn out to be negligible. Much like their no measurement error counterparts in \eqref{eq:barf} and \eqref{eq:hatf}, they will be asymptotically gaussian.
\end{comment}
Let us first show that $\overline{F^*}$ is a biased estimator of $\mu_F$. To this aim, let us apply the Central Limit Theorem to perform the following computations. 
    \begin{equation}
        \label{eq:fstarbarasymp}
        \sqrt{n}\left(\overline{F^*}-\mathbb{E}(F^*)\right) \to \mathcal{N} \left(0,  \sigma^2_{F^*}\right),
    \end{equation}
Furthermore, by linearity
\begin{align*}
    \mathbb{E(}\overline{F^*})&=\mathbb{E}(F^*)\\
    &=\mu_F +\frac{Cov(F^*, X^*)}{\mathbb{E}(1-X)}
\end{align*}
Consequently, $\overline{F^*}$ is a biased estimator of $\mu_F$, having a bias of $\dfrac{Cov(F^*,X^*)}{1-\mu_X}$, regardless of the sample size $n$. Furthermore, observe that as both $1-X^*$ and $\epsilon_Y$ are independent of $F$, one has:

\begin{align}
 \label{eq:f_star}
F^*&=\frac{Y^*-1}{1-X^*}+1 \text{, and thus, } \\
\mathbb{E}(F^*) &=\mathbb{E}\left(\mathbb{E}(\frac{Y^*-1}{1-X^*}|F)\right)+1 \\
 &=(\mu_Y-1)\mathbb{E}((1-X^*)^{-1})+1
\end{align}
 
 the next computations use Jensen's inequality in the second line and show that the bias is always zero or negative: 

\begin{align}
    \frac{Cov(F^*, X^*)}{\mathbb{E}(1-X)}&=(\mu_Y-1)\mathbb{E}((1-X^*)^{-1})+1-\mu_F\\ %no entiendo
    &\leq(\mu_Y-1)[\mathbb{E}(1-X^*)]^{-1}+1-\mu_F\\
    &\leq \frac{\mu_Y-1}{1-\mu_X}+1-\mu_F\\
    &\leq 0 \label{ref:covlesszero}
\end{align}

To show that $\widehat{F^*}$ is an asymptotically unbiased estimator of $\mu_F$, let us notice that the errors are zero-mean and uncorrelated. Hence, the Central Limit Theorem guarantees the vector of observed average pretest and post-test scores satisfies

   \begin{equation}
        \label{eq:vectorasymp}
\sqrt{n}\left(\begin{pmatrix}\overline{X^*} \\ \overline{Y^*} \end{pmatrix} - \begin{pmatrix}\mu_X\\ \mu_Y \end{pmatrix}\right) \to \mathcal{N} \left(\begin{pmatrix}0 \\ 0 \end{pmatrix},  \Sigma + \begin{bmatrix}\delta^2_X & 0 \\ 0 & \delta^2_Y \end{bmatrix}\right).
    \end{equation}

    Where $\Sigma$ is the covariance matrix of the vector of scores $\begin{pmatrix}X \\ Y \end{pmatrix}$, $\delta^2_X$ and $\delta^2_Y$ are the variances of the errors $(\epsilon_{X}, \epsilon_{Y})$. The function $f(x,y)=\frac{y-x}{1-x}$ is continuously differentiable on $0<x<y<1$, the domain of theoretical scores $(X,Y)$ with its gradient given by $\nabla f(x,y)=\begin{pmatrix}
        \frac{y-1}{(1-x)^2} \\ \frac{1}{1-x} \end{pmatrix}$ and, furthermore, $f(\mathbb{E}(X), \mathbb{E}(Y)) =\frac{\mathbb{E}(F(1-X))}{\mathbb{E}(1-X)}=\mathbb{E}(F)$. By the delta method \cite{Shao_2003_book}, one has

         \begin{equation}
        \label{eq:fstarhatasymp}
        \sqrt{n}\left(\widehat{F^*}-\mathbb{E}(F)\right) \to \mathcal{N} \left(0,  \sigma^2_{\widehat{F*}}\right).
    \end{equation}

Where $\sigma^2_{\widehat{F^*}}=[\left(\mathbb{E}(Y)-1\right)^2 (\sigma^2_X+\delta_X^2) + 2\sigma^2_X(\mathbb{E}(Y)-1)(\mathbb{E}(Y)-\mathbb{E}(X)) +(\sigma^2_Y+\delta_Y^2)(1-\mathbb{E}(X))^2](1-\mathbb{E}(X))^{-4} <\infty $ as $\mathbb{E}(X) < 1$. By the Chebyshev inequality\cite{Bill86} one has 

\begin{equation}
    \label{eq:pconvfhatstar}
    \mathbb{P}(|\widehat{F^*}-\mathbb{E}(F)|>\varepsilon) \leq \frac{2\sigma^2_{\widehat{F}}}{n\varepsilon^2} 
\end{equation}

For large enough $n$ and arbitrary $\varepsilon>0$ thanks to \eqref{eq:fstarhatasymp}, and thus, as the sample size $n$ grows large, by \eqref{eq:pconvfhatstar}, $\widehat{F^*}$ converges to $\mu_F$ in probability. This implies that $\widehat{F^*}$ is an asymptotically unbiased estimator of $\mu_F$. Consequently, 
\[\mathbb{E}(\widehat{F^*}) = \mu_F\]

In summary, we have shown that $\widehat{F^*}$ is an asymptotically unbiased estimator of $\mu_F$ whereas $\overline{F^*}$ is biased. Additionally, we have shown that:
\begin{align*}
    \mathbb{E}(\overline{F^*})&=\mu_F +\frac{Cov(F^*, X^*)}{\mathbb{E}(1-X)}\\
    &=\mathbb{E}(\widehat{F^*})+\frac{Cov(F^*, X^*)}{\mathbb{E}(1-X)} +o(n^{-1})
\end{align*}
where $o(n^{-1})$ i.e. that $\lim_{n\to \infty }o(n^{-1}) \to 0$, and thus, it denotes a quantity that decays to $0$ faster than $\frac{1}{n}$. Consequently, when considering large enough $n$, we can compute the difference between the two estimators as follows:

\begin{align*}
    \mathbb{E}(\overline{F^*})-\mathbb{E}(\widehat{F^*})
    & \to \frac{Cov(F^*, X^*)}{\mathbb{E}(1-X)}, \text{ and then, taking a limit} \\
        &= \frac{\mathlarger{\rho}_{F^* X^*}}{\mathbb{E}(1-X)} \sigma_{F^*} \sigma_{X^*}
\end{align*}
Hence, when $n$ is large enough, the difference between the two estimators' means has the same sign as the correlation between the noisy measures of the pretest $(X^*)$ and ngains $(F^*)$. More precisely, we obtained a result that is consistent with Bao's statement (\ref{baoStatement}) previously discussed in prior research \cite{bao_theoretical_2006}, namely that:

\begin{align}
\label{thisStudyStatement}
\mathbb{E}[\overline{F^*}]- \mathbb{E}[\widehat{F^*}] \geq 0 \textbf{ \ if and only if   \  } \mathlarger{\rho}_{F^* X^*} \geq 0
\end{align}

In summary, the random variable $\widehat{F^*}$ is an asymptotically unbiased estimator for $\mu_F$, whereas the random variable $\overline{F^*}$ is a biased estimator for $\mu_F$, whose bias is negative (see eq. \ref{ref:covlesszero}). Consequently, the estimator $\overline{F^*}$ consistently underestimates $\mu_F$, and thus, generally:
\begin{align}
\mathbb{E}[\overline{F^*}] \leq \mathbb{E}[\widehat{F^*}] = \mathbb{E}[F]
\end{align}

\section{Computational Simulations}\label{sec:comsim}

The results presented in the previous sections will be put under testing in this section. We performed simulations to compare the performance of $\overline{F}$ and $\widehat{F}$ while illustrating the theoretical results regarding the effect of error measurements on the correlation between pretest and ngains. 

\begin{figure}[h!]
	\begin{centering}
\includegraphics[width=\linewidth]{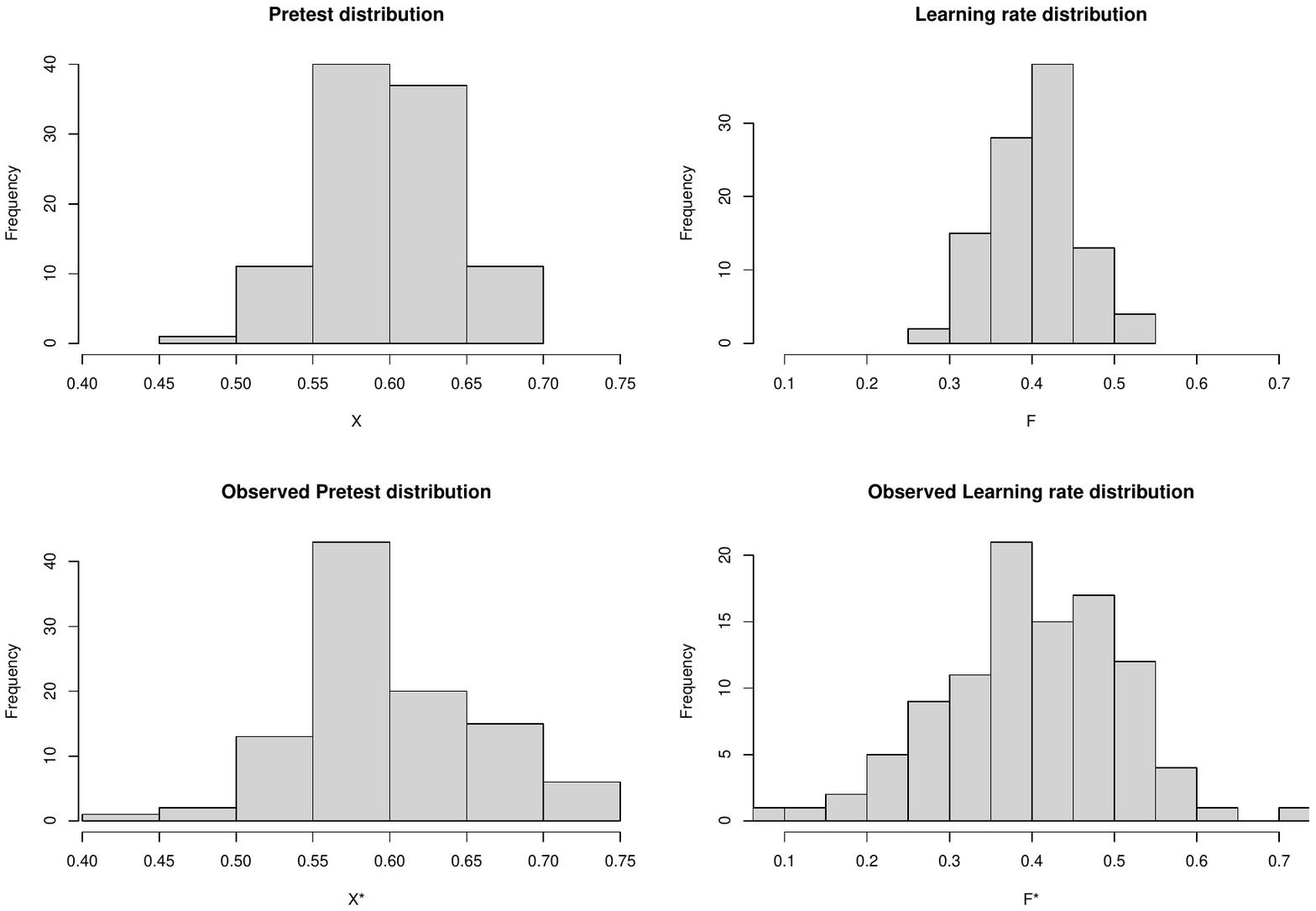}
		\caption{ Simulation of $X$, $F$, $X^*$, $F^*$  in one sample of 100 students.}
		\label{fig:onesample}
	\end{centering}
\end{figure}

First, the sample size was set to $n=100$, to represent common sample sizes in educational practice. We simulate true $X$ and $F$ as distributions in the interval $[0,1]$ as $X \sim Beta(\alpha_x,\beta_x)$ and $F \sim Beta(\alpha_F,\beta_F)$. These distributions are approximately Gaussian for large parameters $\alpha_x$, $\beta_x$, $\alpha_F$ and $\beta_F$ \cite{wise1960normalizing}. Parameters were thus chosen to have bell-shaped empirical distributions $\alpha_x=60$, $\beta_x=40$, $\alpha_F=40$ and $\beta_F=60$, similar to what is seen in real educational data\cite{navarrete_learning_2024}. Using different $X^*$ reliability levels varying between $0.7$ and $1.0$ (a range considered acceptable in educational studies\cite{shultz2020measurement}), for each of them, we defined two vectors of independent Gaussian errors $\epsilon_X\sim N(0,\sigma)$ and $\epsilon_Y\sim N(0,\sigma)$ and, with them, we set
\begin{align*}
X^*&=X+\epsilon_X\\
Y^*&=X+F(1-X)+\epsilon_Y
\end{align*}
subsequently calculating  $F^*$ using the formula from equation \eqref{eq:f_star}. Let us recall that the reliability of $X^*$ is $\rho_{X^*}=\frac{var(X)}{var(X^*)}=1-\frac{\sigma}{var(X^*)}$. Figure \ref{fig:onesample} presents the distribution of $X$, $F$, $X^*$, $F^*$ in a single sample of 100 students with $\rho_{X^*}=0.7$, to illustrate the effect of high levels of measurement errors in the observed scores.

We repeated the procedure in the previous paragraph $N=1000$ times to obtain estimates of $\overline{F}$ and $\widehat{F}$ across all reliability levels. The choice of varying noise levels (reliability) while keeping the same $X$ and $F$ data across each of the $1000$ simulations allows us to illustrate the effect of the noise in the estimates, particularly $\overline{F}$.

We illustrate the different performance of the two proposed estimators in our simulations in Figure \ref{fig:dist} below. Plot A shows that the measurement error of $X^*$, quantified trough the reliability, induces a negative correlation between $F^*$ and $X^*$. Theoretically, $F$ and $X$ are independent and kept constant as reliability varies in each simulation. Hence, the correlation is not related to the learning phenomenon. Plot B illustrates that the correlation between $F^*$ and $X^*$ produces a systematic bias in $\overline{F}$ but not in $\widehat{F}$ (which remains around the 0 benchmark in the dotted line), consistent with our result presented in equation \eqref{thisStudyStatement}. Plot C  shows that the bias estimate for $\overline{F}$ is a decreasing function of reliability, which is not the case for  $\widehat{F}$ that stays put around the true value of $\mathbb{E}(F)=0.4$.

\begin{figure}[h!]
	\begin{centering}
\includegraphics[width=\linewidth]{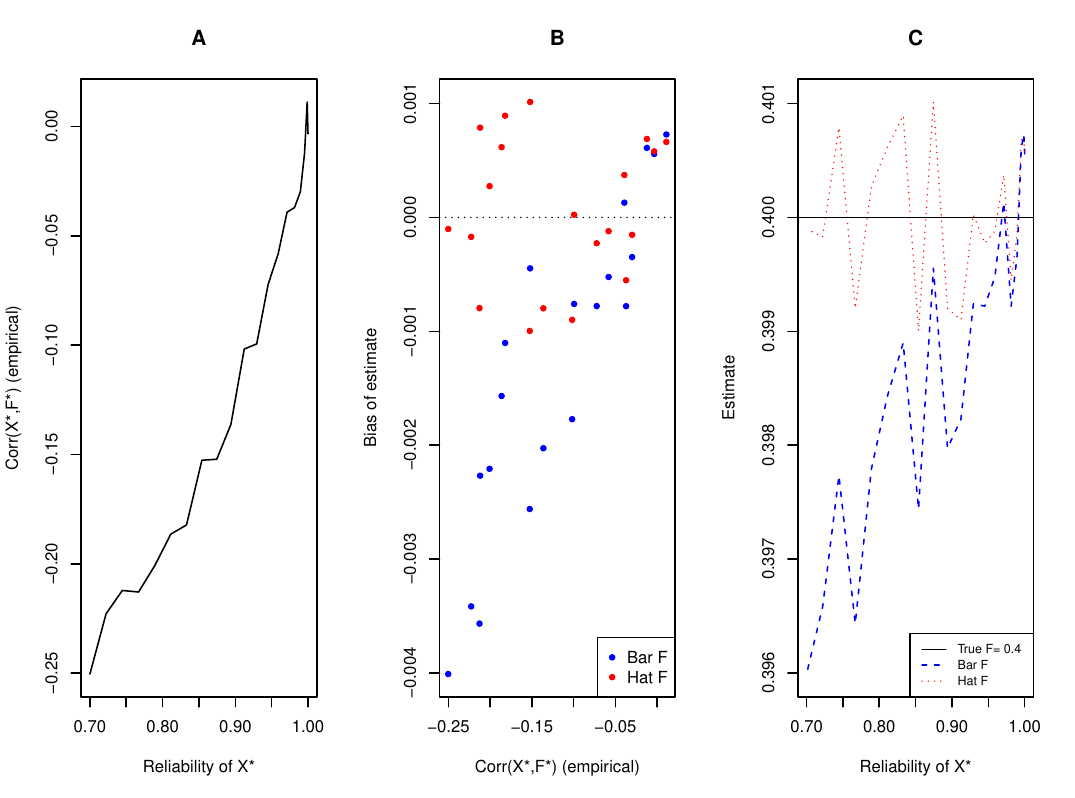}

		\caption{(A) Correlation of observed pretest and observed learning rate as a function of reliability, (B) empirical bias of estimate as a function of observed correlation, and (C) mean value of estimates $\overline{F}$ and $\hat{F}$ as functions of reliability; obtained from synthetic data using $N=1000$ simulations, each with $n=100$ students with beta distributed pretest and learning rate and Gaussian noise. $X\sim\text{Beta}(60,40); F\sim\text{Beta}(40,60)$. 
		}
		\label{fig:dist}
	\end{centering}
\end{figure}

In line with our results in section \ref{sec:noerror}, the case when there are no measurement errors ($\rho_{X^*}=1$ in plots A and C, Fig. \ref{fig:dist}) show no bias in $\overline{F^*}$ neither $\bar{F^*}$. In other cases, a systematic bias can be observed for the $\overline{F}$ distribution, as it consistently under-performs $\hat{F}$ while reliability decreases. Furthermore, reliability shows an almost linear association with correlation between pretest score $X^*$ and learning rate $F^*$, supporting our claim that this covariance is negative and a decreasing function of reliability. This presents an alternative interpretation for the phenomena described by \eqref{baoStatement} in \cite{bao_theoretical_2006} as the difference between estimators is due to measurement errors, which induce a spurious ngains-pretest correlation as our generated distributions of $X$ and $F$ are kept independent throughout our simulations (code available upon request).

\section{Discussion and Conclusions}
According to recent research, ngains estimate a construct called learning rate that describes an ability to acquire unfamiliar knowledge under particular learning conditions \cite{navarrete_learning_2024}. The crucial question here was how to estimate the learning rate of students from pretest-posttest data. The two estimator proposals are $\overline{ng}$ and $\widehat{ng}$ and the discrepancy between them has been proposed to give insight into the learning process of students \cite{bao_theoretical_2006} in the form of a pretest-ngain correlation. Crucially, such a correlation might indicate that ngains---as a measurement of learning---cannot control for the influence of prior knowledge. By assuming the hypothesis of the LR-model, namely that student's prior knowledge and learning rate are independent, this study shows that the difference between the two estimators is due to measurement errors in the pretest and posttest scores (or by randomness). When there are no measurement errors, these two estimators are unbiased; thus, any difference between them is random. When there are measurement errors, $\widehat{ng}$ is an asymptotically unbiased estimator for the learning rate, whereas $\overline{ng}$ is biased. In more detail, the Law of Large Numbers and the Central Limit Theorem indicate that $\widehat{F^*}$ has small errors because $\overline{X^*}$ and $\overline{Y^*}$ have errors whose variances are converging to zero. In the case of $\overline{F^*}$, each observation contributes non-linearly to its error, meaning that the involved errors do not cancel themselves but aggregate systematically. Furthermore, even when the LR-Model assumed independence between the pretest $(X)$ and the learning rate $(F)$, these error contributions induced a spurious negative correlation between the noisy versions of the pretest $(X^*)$ and the learning rate $(F^*)$. Furthermore, we have shown that such a pretest-ngain correlation determines the difference $\overline{ng} - \widehat{ng}$. Consequently, neither such a difference nor the pretest-ngain correlation should be taken as informative about the learning process. Observing a pretest-ngain correlation \emph{does not} imply that the ngain transformation (estimating a learning rate) is biased to favor subgroups of learners according to their pretest performance. 

Additionally, we showed that the two estimators $\widehat{F^*}$ and  $\overline{F^*}$ converge to different values and that 
\begin{align}
%\label{thisStudyStatement} hay que comentar esto porque el label esta antes, y genera error ponerlo aquí
\mathbb{E}[\overline{F^*}]- \mathbb{E}[\widehat{F^*}] \geq 0 \textbf{ \ if and only if   \  } \mathlarger{\rho}_{F^* X^*} \geq 0
\end{align}
which is analogous to a key conclusion of previous research \cite{bao_theoretical_2006} (see statement \ref{baoStatement}). Hence, there is a high level of consistency between the results of this study and previous findings in the literature. Nevertheless, there is a sharp contrast in terms of interpretation. According to Bao's study, the difference between both estimators gives insight into the learner's learning process; more precisely, such a difference quantifies the influence of the pretest on the learning rate and, thus, points out a weakness of ngains as an estimator of learning (rates). In contrast, this study inferred the statement (\ref{thisStudyStatement}) from the LR-Model, which assumes independence between the pretest $(X)$ and the learning rate $(F)$. Consequently, even when the pretest $X$ and the learning rate $F$ are independent, we observed a spurious correlation between the pretest ($X^*$) and ngains ($F^*$) due to the structure of measurement errors. In other words, observing such a pretest-ngain correlation does not preclude that the latent variables, namely the prior knowledge $(X)$ and the learning rate $(F)$ are statistically independent. 

Regarding the usefulness of our results for analyses of learning data, we emphasize that measurement errors are unavoidable in practice, and thus, we only refer to such a case here. On the one hand, $\widehat{ng}$ is asymptotically unbiased and precise as an estimator of a group's learning rate. Hence, at first sight, it seems the optimal estimator for the learning rate. Nevertheless, it is worth noticing that it cannot integrate individual information into data analysis. For example, $\widehat{ng}$ cannot be the output variable of an ANCOVA (or ANOVA) that considers individual information as predictors. In contrast, although $\overline{ng}$ underestimates the learning rate, it can input individual information into data analysis (e.g., cognitive abilities, SES, teaching methods, etc.) to account for or control. For example, prior research has used individual measures of general cognitive abilities to discount its effect when estimating the learning rate of math \cite{navarrete_learning_2024}. Similarly, research in physics education has recommended using Lawson Test scores when estimating learning rates from data associated with the Force Concept Inventory \cite{coletta_interpreting_2007}. In summary, both estimators have strengths and weaknesses. The best of both worlds would be to use $\overline{ng}$ as the output variable of an analysis (e.g., ANCOVA) but then force the computations to center the mean of the output on the more accurate estimate $\widehat{ng}$. Although developing this approach requires further research, it would enjoy the accuracy of an unbiased estimator and the ability to input individual information into the data analysis. Future research might devise computational methods to perform such statistical analysis by correcting the estimations of learning rates by considering the pretest-ngain correlation induced by measurement errors.

It is worth discussing here a particular concern regarding ngains in prior research. Nissen J. et al. have argued ``that [ngains] is biased in favor of high pretest populations''\cite{nissen_comparison_2018} because they have observed pretest-ngains correlations, thus implying ngains favor students with higher prior knowledge. A critical concern in the measurement literature is the possibility that knowledge change (or growth) metrics are biased for or against different groups of students. Hence, the authors recommended using Cohen's d rather than ngains as a measure of learning. These recommendations ignited a heated debate, including a response pointing out a problem with the analysis, namely, the omission of control variables \cite{coletta_why_2020} such as science reasoning abilities that can be measured with Lawson's Test. Our results provide a different alternative: measurement errors in pretest and posttest scores induce a (spurious) correlation between estimates of learning rates (ngains) and pretest scores (pretest). Consequently, despite observing correlations between ngains and pretest scores, ngains estimations may be unbiased regarding groups with different levels of pretest performance. In other words, ngains-pretest correlations can be justified regarding measurement errors.
   
In alignment with Classical Test Theory (CTT), this study included additive errors in both pre-test and post-test observed scores, a widely accepted framework for educational measurement analysis\cite{shultz2020measurement}. However, this approach has two limitations. In the first place, we considered $X$ and $F$ bounded between 0 and 1, and thus, the additive Gaussian errors could take these variables outside of their bounds, which was an aspect that we had to control in the simulations when the mean of $X$ or $F$ is close to either boundary; limiting the possibility of simulating too low or too high pre-test scores or learning rates. The second limitation is that CTT faces criticism from Item Response Theory (IRT) proponents, which offers an alternative. In this regard, IRT models \cite{cai2016item} relate observed scores in a test to a latent variable representing the true score. For dichotomous items, it will use Generalized Linear Models such as Logit or Probit models \cite{aldrich1984linear} to predict the response of each item using the true score. These models add linear measurement errors at the link function scale and not at the response scale \cite{ai2003interaction}, making them mathematically robust at the boundaries and more realistic in psychometry. Future work should extend the framework developed in this study to scores measured according to Item Response Theory models.

\appendix*   % Omit the * if there's more than one appendix.

%\begin{sideways}
\begin{table}[h]
\small

\caption{Some notational changes from prior literature.}
\label{tabla:ejemplo}
\begin{adjustbox}{angle=90,margin=-4cm 0cm 0cm 0cm}
\begin{tabular}{@{}lccccccc@{}} % l (alineación izquierda), c (centrado), r (alineación derecha)
\toprule
\textbf{Description} & \multicolumn{2}{c}{\textbf{Bao's Model}} & \multicolumn{2}{c}{\parbox{3cm}{\textbf{Model without measurement error}}} & \multicolumn{2}{c}{\parbox{3cm}{\textbf{Model with measurement error}}} \\
%\cmidrule(lr){2-3} \cmidrule(lr){4-5} \cmidrule(lr){6-7}
 & R.V. & Realization & R.V. & Realization & R.V. & Realization \\
%\midrule % NO SE PORQUE YA NO FUNCIONA :S SORRY MARIANO AYUDA (SOY LA VALE)
Pretest          &  $X$   & $x$                         & $X$  & $x$& $X^*=X+\varepsilon_X$ & $x^*$ \\
Postest          & $Y$ & $y$                         &   $Y$     & $y$ & $Y^*=Y+\varepsilon_X$ & $y^*$ \\
Average pretest  & $\overline{X}=\frac{1}{N}\sum X_i$& $\overline{x}$              & $\overline{X}=\frac{1}{N}\sum X_i$& $\overline{x}$ & $\overline{X^*}=\frac{1}{N}\sum X^*_{i}$ & $\overline{x^*}$ \\
Average postest  & $\overline{Y}=\frac{1}{N}\sum Y_i$ & $\overline{y}$              & $\overline{Y}=\frac{1}{N}\sum Y_i$ & $\overline{y}$ & $\overline{Y^*}=\frac{1}{N}\sum Y^*_{i}$ & $\overline{y^*}$ \\
Gain             & & $\overline{y}-\overline{x}$ & $G=Y-X$ & $g=y-x$ & $G^*=Y^*-X^*$ & $g^*$ \\
Normal gain     & $G=\frac{Y-X}{1-X}$ & $g=\frac{y-x}{1-x}$ & $F=\dfrac{Y-X}{1-X}$ & $f=\dfrac{y-x}{1-x}$ & $F^*=\frac{Y^*-X^*}{1-X^*}$ & $f^*=\frac{y^*-x^*}{1-x^*}$  \\
Case Average & $G=\frac{\overline{Y}-\overline{X}}{1-\overline{X}}$ & $g=\frac{\overline{y}-\overline{x}}{1-\overline{x}}$              & $\hat{F}=\dfrac{\overline{Y}-\overline{X}}{1-\overline{X}}$ & $\hat{f}=\dfrac{\overline{y}-\overline{x}}{1-\overline{x}}$ & $\hat{F^*}=\dfrac{\overline{Y^*}-\overline{X^*}}{1-\overline{X^*}}$ & $\hat{f^*}=\dfrac{\overline{y^*}-\overline{x^*}}{1-\overline{x^*}}$ \\
Average of cases & $\overline{G}=\frac{1}{N}\cdot \sum_{k=1}^\infty G_k$ & $\overline{g}=\frac{1}{N}\cdot \sum_{k=1}^\infty g_k$             & $\overline{F}=\frac{1}{N}\sum_{i=1}^N F_i$ & $\overline{f}=\frac{1}{N}\sum_{i=1}^N f_i$ & $\overline{F^*}=\frac{1}{N}\sum_{i=1}^N F^*_{i}$ & $\overline{f^*}=\frac{1}{N}\sum_{i=1}^N f^*_{i}$ \\

%\bottomrule
\end{tabular}
\end{adjustbox}
\end{table}
%\end{sideways}

\section{Further Simulations}

This section contains further simulations following the same procedure as that conducted in section \ref{sec:comsim} but for different parameter settings of pre-test $X$ and learning rate $F$, to better illustrate our results in a wider variety of settings. 

\begin{figure}[h!]
	\begin{centering}
\includegraphics[width=0.8\linewidth]{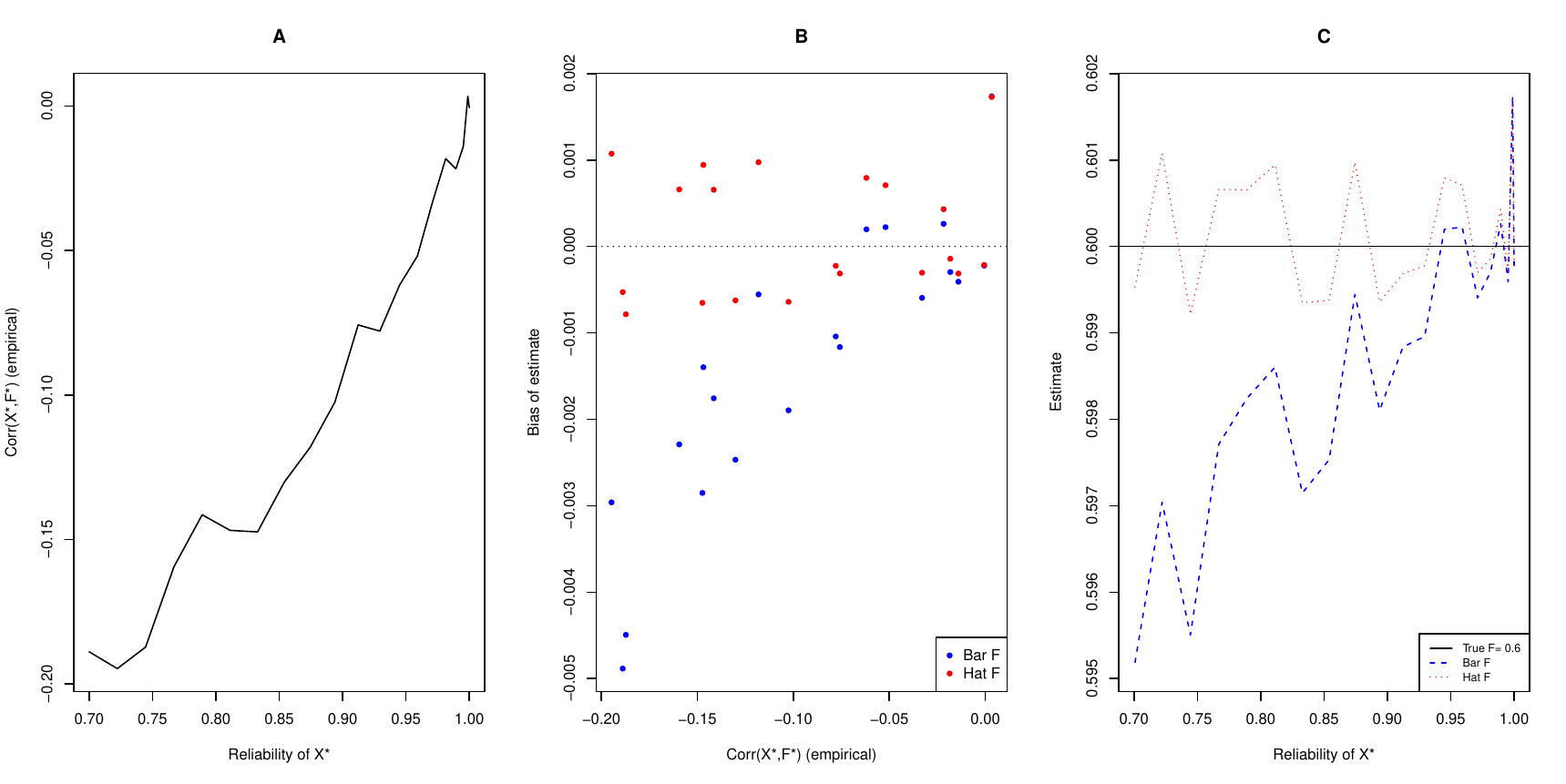}

		\caption{A) Correlation of observed pretest and observed learning rate as a function of reliability, B) empirical bias of estimate as function of observed correlation, and C) mean value of estimates $\overline{F}$ and $\hat{F}$ as functions of reliability; obtained from synthetic data using $N=1000$ simulations, each with $n=100$ students with beta distributed pre test and learning rate and Gaussian noise. $X\sim\text{Beta}(70,30); F\sim\text{Beta}(60,40)$. 
		}
		\label{fig:app1}
	\end{centering}
\end{figure}

\begin{figure}[h!]
	\begin{centering}
\includegraphics[width=0.8\linewidth]{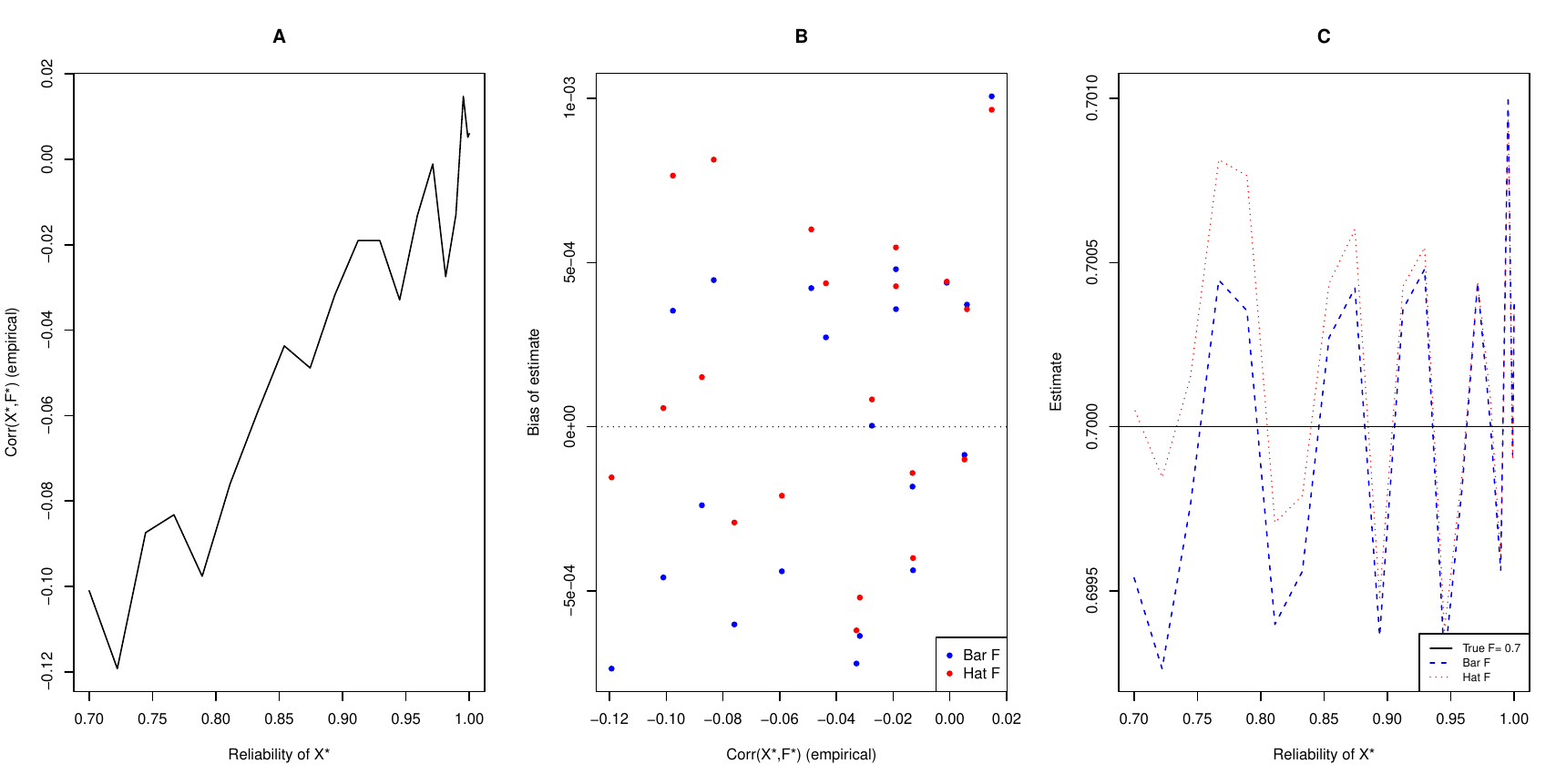}

		\caption{A) Correlation of observed pretest and observed learning rate as a function of reliability, B) empirical bias of estimate as function of observed correlation, and C) mean value of estimates $\overline{F}$ and $\hat{F}$ as functions of reliability; obtained from synthetic data using $N=1000$ simulations, each with $n=100$ students with beta distributed pre test and learning rate and Gaussian noise. $X\sim\text{Beta}(30,70); F\sim\text{Beta}(70,30)$. 
		}
		\label{fig:app2}
	\end{centering}
\end{figure}

\begin{figure}[h!]
	\begin{centering}
\includegraphics[width=0.8\linewidth]{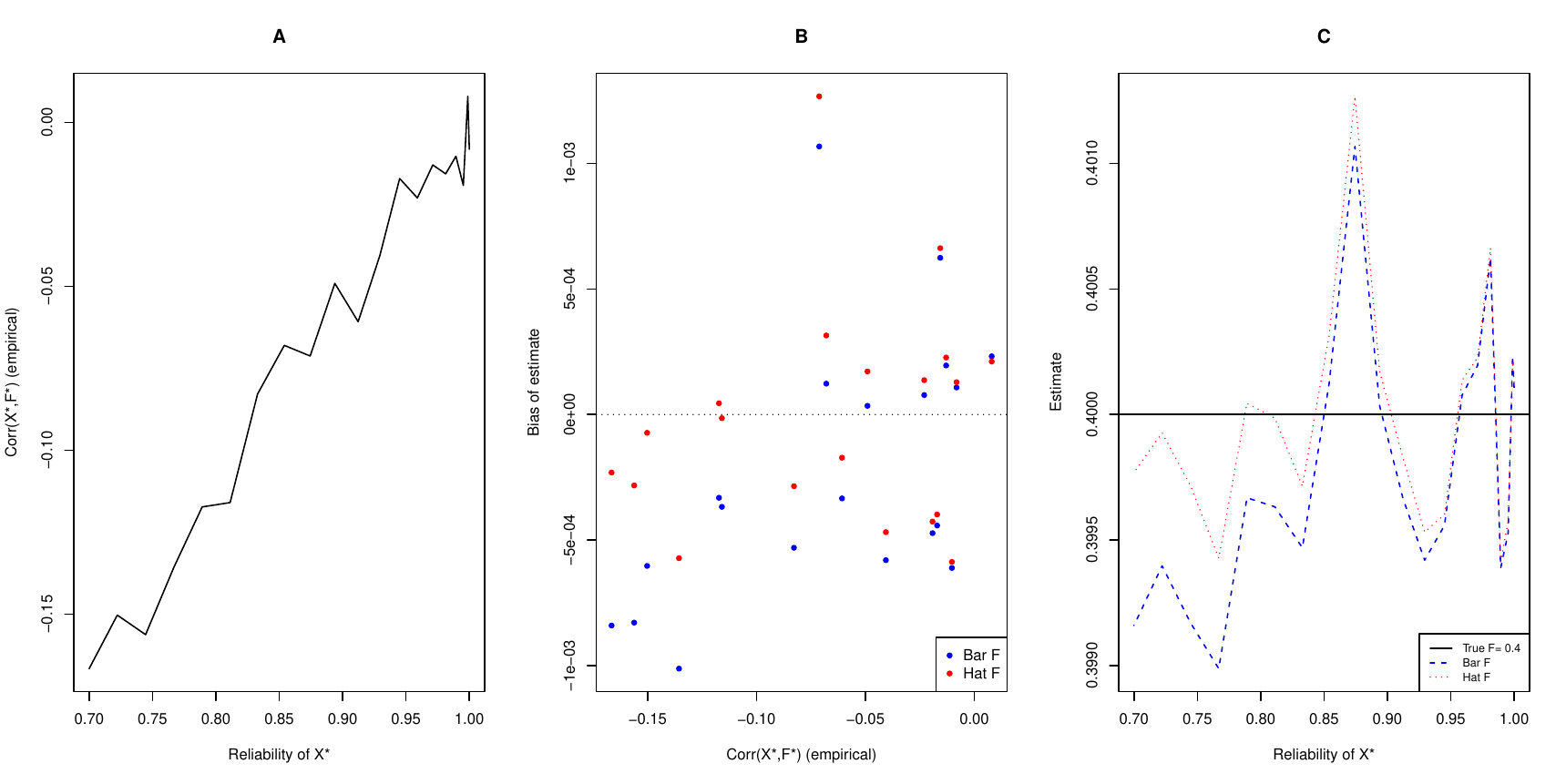}

		\caption{A) Correlation of observed pretest and observed learning rate as a function of reliability, B) empirical bias of estimate as function of observed correlation, and C) mean value of estimates $\overline{F}$ and $\hat{F}$ as functions of reliability; obtained from synthetic data using $N=1000$ simulations, each with $n=100$ students with beta distributed pre test and learning rate and Gaussian noise. $X\sim\text{Beta}(20,80); F\sim\text{Beta}(40,60)$. 
		}
		\label{fig:app3}
	\end{centering}
\end{figure}

\begin{figure}[h!]
	\begin{centering}
\includegraphics[width=0.8\linewidth]{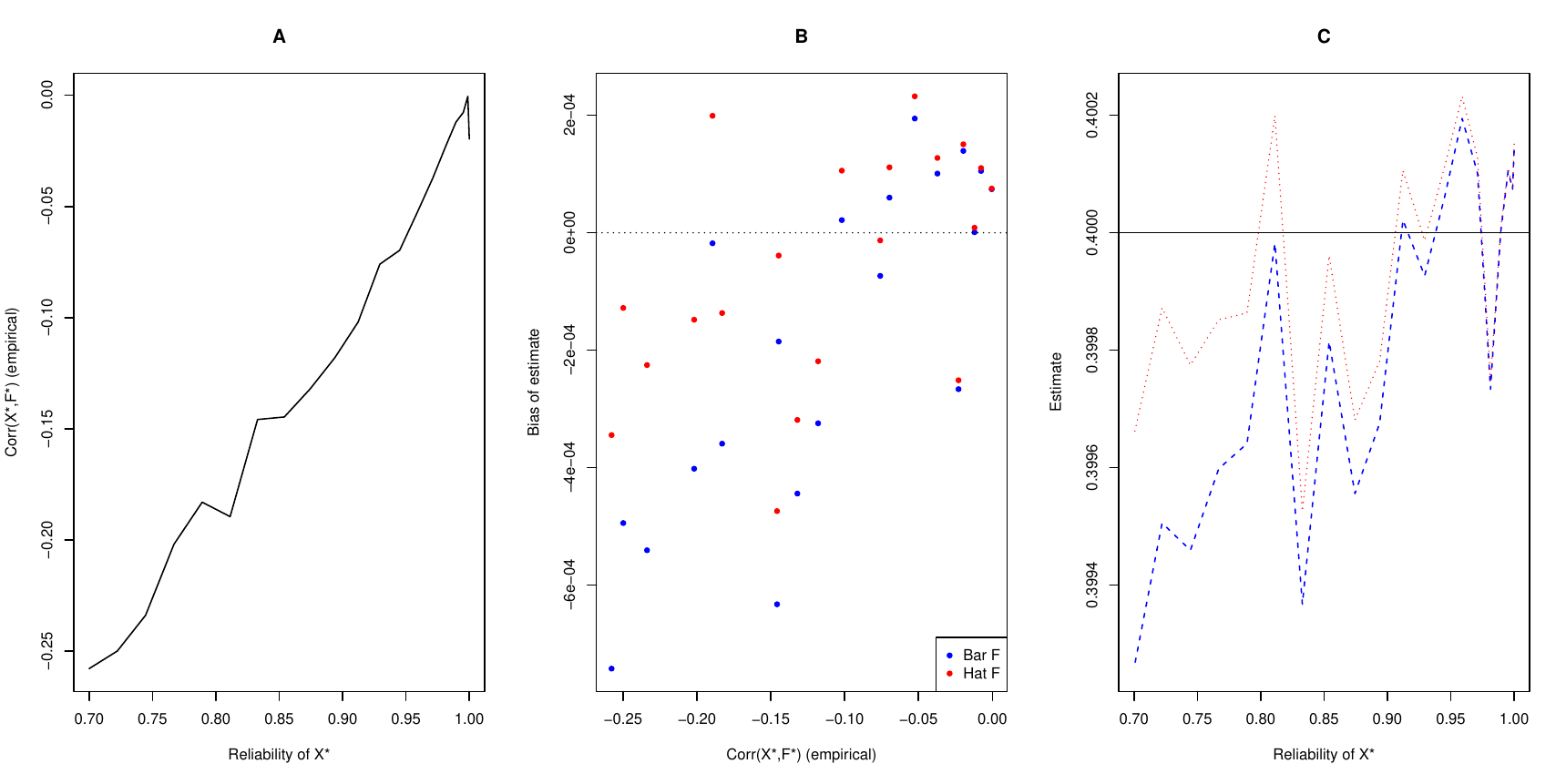}

		\caption{A) Correlation of observed pretest and observed learning rate as a function of reliability, B) empirical bias of estimate as function of observed correlation, and C) mean value of estimates $\overline{F}$ and $\hat{F}$ as functions of reliability; obtained from synthetic data using $N=1000$ simulations, each with $n=100$ students with beta distributed pre test and learning rate and Gaussian noise. $X\sim\text{Beta}(600,400); F\sim\text{Beta}(400,600)$. 
		}
		\label{fig:app4}
	\end{centering}
\end{figure}

Figure \ref{fig:app1} is based on a theoretical pre-test and learning rate  mean of 0.3 and 0.7, respectively and shows a similar pattern than that observed in \ref{fig:dist}. Figures \ref{fig:app2} and \ref{fig:app3} are based on theoretical pre-tests means of 0.3 and 0.2, and learning rate means of  and 0.7, respectively; and they show that in the low pre test case $\overline{F}$ is highly correlated to $\hat{F}$ (plots C) and that its bias is less noticeable (plot B), suggesting that the mean-of-cases estimator may be reliable in this scenario. Figure \ref{fig:app4} is based on  theoretical pre-test and learning rate  mean of 0.6 and 0.4, as figure \ref{fig:dist} is, but the beta parameters are all multiplied by 10 to reduce the variance of both variables. In this case, we observe a much smaller bias of $\overline{F}$ when compared to the scenario simulated in section \ref{sec:comsim} that is, nevertheless, still noticeably decreasing on reliability. In all simulations, we observe a linear relationship between reliability of $X^*$ and the empirical covariance between $X^*$ and $F^*$, further strengthening the argument that the relationship observed in \cite{bao_theoretical_2006} can be attributed to measurement errors.


\newpage
\begin{acknowledgments}
The funding for this project was thanks to grants ANID/FONDEF/IT23I0012 granted to J.N. and ANID/PIA/Basal Funds for Centres of Excellence FB210005 and FB0003 that supported V.G.
\end{acknowledgments}

\bibliography{SampleAnonymous.bib}
\end{document}